\documentclass[12pt]{article}

\usepackage{amsmath,amssymb,amsthm}

\newcommand{\EQUA}{\begin{equation}}
\newcommand{\EQQN}{\end{equation}}
\newcommand{\EQN}{\begin{eqnarray}}
\newcommand{\ENN}{\end{eqnarray}}
\newcommand{\CR}{\nonumber \\}

\newcommand{\EX}{{\rm e}}
\newcommand{\I}{i}%{{\rm i}}
\newcommand{\PA}{\partial}

\newcommand{\B}{\beta}

\newcommand{\G}{\gamma}
\newcommand{\GA}{\varGamma}

\newcommand{\E}{\epsilon}
\newcommand{\VE}{\varepsilon}

\newcommand{\HA}{{1 \over 2}}

\begin{document}

\title{\Large\bf Statistical Mechanics for Unstable States \\
in  Gel'fand Triplets 
and \\ Investigations of Parabolic Potential Barriers}

\author{ Tsunehiro Kobayashi\footnote{E-mail: 
kobayash@a.tsukuba-tech.ac.jp} 
and Toshiki Shimbori\footnote{E-mail: 
shimbori@het.ph.tsukuba.ac.jp} \\
{\footnotesize\it $^*$Department of General Education 
for the Hearing Impaired,}
{\footnotesize\it Tsukuba College of Technology} \\
{\footnotesize\it Ibaraki 305-0005, Japan}\\
{\footnotesize\it $^\dag$Institute of Physics, University of Tsukuba}\\
{\footnotesize\it Ibaraki 305-8571, Japan}}

\date{}

\maketitle

\begin{abstract}
Free energies and other thermodynamical quantities are investigated 
in canonical and grand canonical ensembles of statistical mechanics 
involving unstable states which are described by the generalized eigenstates 
with complex energy eigenvalues in  
the conjugate space of Gel'fand triplet. 
The theory is applied to the systems containing 
parabolic potential barriers (PPB's). 
The entropy and energy productions from PPB systems are studied. 
An equilibrium for a chemical process described by reactions 
$A+CB\rightleftarrows AC+B$ is also discussed.
\end{abstract}
\thispagestyle{empty}

\setcounter{page}{0}

\pagebreak
\hfil\break
{\bf 1. Introduction}
\vskip5pt

Many experimental and theoretical investigations show that 
thermodynamics is a fundamental dynamics for describing realistic phenomena 
governed by temperatures. 
We also know that quantum mechanics is a fundamental one 
to describe microscopic processes. 
And we believe that statistical mechanics is a theory connecting 
quantum mechanics to thermodynamics. 
In statistical mechanics we know that 
so-called ``principle of equal {\it a priori} probability'' 
is taken as the guiding principle in the construction of the theory 
and 
the Boltzmann entropy is the key word connecting the 
two fundamental dynamics. 
Rigorously speaking, thermodynamics is applicable only to 
true equilibriums described by the maximums of entropies. 
We, however, know the fact that thermodynamics is applicable to 
phenomena which are slowly varying with time, such as phenomena in chemical 
processes, cosmological processes and so on. 
This fact indicates 
that the principle of thermodynamics can also be applicable 
to those phenomena varying very slowly as compared with 
time-scales needed for making thermal equilibriums locally. 
In statistical mechanics states included in the count of thermodynamical 
weights are the eigenstates of quantum mechanics which can have only 
real energy eigenvalues on Hilbert spaces. 
All eigenstates in Hilbert spaces are stable and then there is no 
possibility for introducing the changes with respect to 
time in statistical mechanics 
based on quantum mechanics on Hilbert spaces. 
At present, therefore, 
we have no reliable theory to investigate paths which connect 
an initial equilibrium to a final equilibrium. 
Taking account of the fact that thermodynamics can be applicable to 
some phenomena slowly varying with time, it seems to be very interesting that 
we examine statistical mechanics on some extended spaces including unstable 
states. 
For this purpose 
we find out an interesting possibility of the extension of 
Hilbert spaces to the conjugate spaces in Gel'fand triplets [1], 
where complex energy eigenvalues describing unstable states 
are involved. 

In the previous paper [2] we have shown the fundamental idea of the extension 
of statistical mechanics 
on Hilbert spaces to that on the conjugate spaces of Gel'fand triplets  
on the basis of principle of equal {\it a priori} probability 
and derived canonical distributions with a common time-scale. 
The fundamental difference between statistical mechanics on Hilbert spaces 
(SMHS) 
and that on Gel'fand triplets (SMGT) appears in the 
count of the states for the evaluation of thermodynamical weight, 
that is,  the new freedom arising from the states with imaginary eigenvalues 
appears in SMGT, while there is no such freedom in SMHS. 
This fact changes the entropy $S$ which is defined by 
\begin{equation}
S({\cal E})=k_{\rm B}\log W({\cal E}), 
\end{equation}
where $W({\cal E})$ is the thermodynamical weight 
at the total complex energy 
${\cal E}=E-i\GA$ and $k_{\rm B}$ is the Boltzmann constant. 
In the evaluation of $W({\cal E})$ 
two freedoms that arise from the variety of the combinations 
for composing the real part of the total energy $E=\sum_i\E_i$ 
and that for composing 
the imaginary one $\GA=\sum_i\G_i$ must be taken into account, 
where $\E_i$ and $\G_i$, respectively, denote the real and imaginary parts 
of the complex energy eigenvalue $\varepsilon_i=\E_i-\I \G_i$ 
for the $i$th constituent. 
Provided that there is no correlation between the real energy eigenvalues and 
the imaginary ones,  $W$ is given by the product of 
the thermodynamical weight for the 
real part $W^\Re(E)$ and that for the imaginary part $W^\Im(\GA)$ 
\begin{equation}
W({\cal E})=W^\Re(E)W^\Im(\GA).
\end{equation}
Thus the entropy in SMGT is represented by the sum of the Boltzmann 
entropy $S^\Re(E)$ and the new one 
$S^\Im(\GA)$ induced from the freedom of the 
imaginary energy eigenvalues such that 
\begin{equation}
S({\cal E})=S^\Re(E)+S^\Im(\GA), 
\end{equation}
where $S^\Re(E)=k_{\rm B}\log W^\Re(E)$ and 
$S^\Im(\GA)=k_{\rm B}\log W^\Im(\GA)$. 
An explicit example for eqs.~(2) and (3) was presented in ref.~3 by using 
parabolic potentials. 
The canonical distribution has also been derived as 
\begin{equation}
P({\cal E}_{lm})=Z^{-1}\exp (-\B^\Re E_l-\B^\Im\GA_m), 
\end{equation}
where the canonical partition function 
is given by 
$$
Z=\sum_l\sum_m \exp (-\B^\Re E_l-\B^\Im\GA_m).
$$  
In the partition function the two $\B$ factors are related to the two 
temperatures as 
\begin{equation}
\B^\Re\equiv \B=(k_{\rm B}T)^{-1},\ \ \ \B^\Im=(k_{\rm B}T^\Im)^{-1},
\end{equation} 
where $T$ is the usual temperature of canonical distributions and 
$T^\Im$ is newly introduced in SMGT [2]. 
Comparing the time-dependence of the probability distributions 
for the quantum states on Gel'fand triplets 
having the total imaginary energy $\GA$, 
which is given by 
$e^{-2\GA t/\hbar}$, 
with that of the canonical 
distribution, we have derived the relation $\B^\Im=2t/\hbar$ 
with the common time-scale $t$ [2], that is, $T^\Im=\hbar/2k_{\rm B}t$. 
(In details, see ref.~2.) 
We should understand that the canonical distribution is meaningful when 
$|\GA|$ is small enough to make a thermal equilibrium before the change of the 
physical properties of the total system. 
In fact we see that such situations can happen, that is, 
$|\GA|$ can be as small as possible, including exact zero value, 
because in Gel'fand triplet formalism [1] 
all eigenvalues appear in the pair of 
complex conjugates such as $\E\mp i\G$  and then 
the total imaginary part $\GA$ can be zero. 
It is a striking fact 
that there exist stable systems  
which are composed of unstable states. 
An example for the stable systems was presented in ref.~2 in terms of 
2-dimensional parabolic potential barriers (PPB's). 
It should also be noted that in the 2-dimensional PPB we can show the 
existence of stationary states with zero imaginary eigenvalue which are 
understood as stationary flows round the center of PPB [4].
By using the stationary states the energy and entropy productions from PPB 
were studied and the entropy transfer from $S^\Im$ to $S^\Re$ was 
suggested [3]. 
This new idea for statistical mechanics 
seems to have many interesting applications such as 
chemical processes, energy production processes without nuclear fusions, 
the birth of the Universe and so forth. 

In the previous paper [2] we presented the fundamental idea for the 
extension of SMHS to SMGT and derived the canonical distribution with the 
common time-scale. 
The presentation is, however, not enough to understand SMGT well, 
for instance, thermodynamical functions except the entropy are not discussed. 
In this paper we would like to investigate the new statistical mechanics 
i.e. SMGT 
involving unstable states on Gel'fand triplets more precisely. 
Namely, thermodynamical quantities such as free energies and 
chemical potentials  will be investigated in SMGT in 
$\S\S$~2 and 3. 
Consistency of the theory will be examined in terms of simple PPB models 
in $\S$~4. 
The entropy transfer from $S^\Im$ to $S^\Re$ and the energy production are 
studied through a decay of a resonance system in PPB  in $\S$~5. 
An equilibrium for a simple process described by reactions 
$A+CB\rightleftarrows AC+B$ 
will be discussed in this scheme in $\S$ 6. 
Throughout this paper we deal with the processes in which the real and the 
imaginary parts of the total energy of the system can be independently 
determined such as the case of parabolic potentials presented in refs.~2 
and 3. 

\vskip15pt
\hfil\break
{\bf 2. Free energies in canonical ensemble}
\vskip5pt

Let us start from the canonical distribution of (4). 
In the present case where the real and the imaginary energies of the 
system can be independently determined, 
the canonical partition function for the system composed of $N$ constituents 
can be obtained as the product of 
the partition function for the real part and that for the imaginary one 
such that 
\begin{equation}
Z_N(T,t)=Z_N^\Re(T) Z_N^\Im(t), 
\end{equation} 
where 
$$
Z_N^\Re(T)=\sum_l \exp (-\B E_l), \ \ \ 
Z_N^\Im(t)=\sum_m \exp (-\B^\Im \GA_m) .  
$$ 
Following the same argument carried out in SMHS, 
we have two (Helmholtz) free energies corresponding to the usual free energy 
for the real part $F^\Re$ and that for the imaginary part $F^\Im$ as 
\begin{equation}
F^\Re(T)=-\B^{-1}\log Z_N^\Re, \ \ \ 
F^\Im(t)=-(\B^\Im)^{-1} \log Z_N^\Im. 
\end{equation}
The mean energies are obtained as usual 
\begin{equation}
\bar E={\partial \over \partial \B}\left[\B F^\Re(T)\right], 
%\log Z_N^\Re, 
\ \ \ 
\bar \GA={\partial \over \partial \B^\Im}\left[\B^\Im F^\Im(T)\right]. 
%\log Z_N^\Im. 
\end{equation}
The relations with respect to other quantities derived from $F^\Re$ 
such as the total volume $V$, the pressure $p$ and so forth
are same as SMHS. 
At present, however, it is not an easy problem to clarify 
whether  new quantities derived from $F^\Im$ are physically 
meaningful or not.  
The entropies $S^\Re$ and $S^\Im$ are derived from the free energies as 
\begin{equation}
S^\Re=-{\partial \over \partial T}F^\Re(T), \ \ \ 
S^\Im=-{\partial \over \partial T^\Im}F^\Im(t).  
\end{equation} 
The consistency of $S^\Im$ given in (9) with that of (3) given 
in microcanonical ensemble [2] will be studied in a PPB model in $\S$ 4. 
In general the entropy $S^\Im$ and the mean value $\bar \GA$ have 
time-dependence, 
which will also be investigated in the PPB model. 
The free energies satisfy the usual relation of SMHS such that 
\begin{equation}
F^\Re=\bar E-TS^\Re,\ \ \ F^\Im=\bar \GA- T^\Im S^\Im. 
\end{equation}
Since we do not know what are good observables in unstable systems 
and still have only one example of PPB to adopt SMGT [3], 
we have to examine SMGT more in other realistic examples %[3,4,5]
in order to understand the meanings of SMGT in details. 
\vskip15pt

\hfil\break
{\bf 3. Grand canonical ensemble}
\vskip5pt

The most prominent aim of SMGT is the introduction of time-dependence 
through the decay of the constituents of systems. 
This means that the total number of constituents composing the systems 
also varies with time. 
This situation will be well described in grand canonical ensemble. 
In the construction of grand canonical ensemble the number of the constituents 
should be represented by natural numbers $N$ ($N=0,1,2,\cdots$). 
Then we construct the grand partition function as 
\begin{equation}
\varXi=\sum_{N=0}^\infty {\rm e}^{\B\mu N}Z_N,
\end{equation}
where $Z_N$ is the partition function for the total number $N$ 
and given by the product $Z_N^\Re Z_N^\Im$. 
In the definition of $\varXi$ the usual factor $\B$ is taken so as to coincide 
with the partition function of SMHS when the freedom of the imaginary part 
disappears. 
The chemical potential $\mu$, of course, differs from that of SMHS and 
generally has the time-dependence. 
The specific difference of $\varXi$ from $Z_N$ is seen in the forms 
of (6) and (11), 
that is, the contributions from the real and the imaginary parts cannot be 
separated in $\varXi$, whereas they are separated as the product in $Z_N$. 
We, therefore, have only one thermodynamical function in the grand 
canonical ensemble given by 
\begin{equation}
J(T,t,\mu)=-\B^{-1}\log \varXi. 
\end{equation}
The mean number is obtained by 
\begin{equation}
\bar N=\B^{-1}{\partial \over \partial \mu}\log \varXi 
\end{equation}
which have the time-dependence in general. 
An example of the time-dependence will be seen in a PPB model. 

Taking into account that the contributions of 
the real and imaginary parts are not separable 
in $\varXi$, 
the maximum of the probability in the grand canonical ensemble appears at  
\begin{equation}
J/T=\bar E/T-S^\Re+\bar \GA/T^\Im -S^\Im-\bar N
\left(\partial S/\partial N\right), %{\partial S \over \partial N}, 
\end{equation}
where the relations 
$ \partial S^\Re/\partial E=1/T$ and 
$\partial S^\Im/\partial \GA=1/T^\Im
$ 
are used [2]. 
Now we can see that the definition of the chemical potential $\mu$ is 
given by the relation 
\begin{equation}
{\mu \over T}=-{\partial S \over \partial N}, 
\end{equation} 
where $S=S^\Re+S^\Im$. 
The Gibbs free energy $G$ is given as usual 
\begin{equation}
G=\mu \bar N. 
\end{equation} 
Note that the relation between the thermodynamical functions  
$J=F-G$ in SMHS should not be adopted. 
In SMGT the relation should be read as 
\begin{equation}
J/T=F^\Re/T+F^\Im/T^\Im -G/T. 
\end{equation} 

In simple cases where 
all constituents can be treated as independent each other, 
the canonical partition function is written by the 
\begin{equation}
Z_N=(Z_1)^N, 
\end{equation} 
where $Z_1=Z_1^\Re Z_1^\Im$ is the partition function for one constituent. 
We then obtain 
\begin{equation}
\varXi=(1- {\rm e}^{\B\mu}Z_1)^{-1} 
\end{equation} 
with the constraint for the chemical potential 
$$
{\rm e}^{\B\mu}Z_1 < 1.
$$ 
When the constituents cannot be identified each other such as free particles, 
we should have 
\begin{equation}
\varXi=\sum_{N=0}^\infty {\rm e}^{\B\mu N}{(Z_1)^N \over N!} 
\end{equation} 
and then we get 
$$
\varXi=\exp ({\rm e}^{\B\mu}Z_1). 
$$ 

\pagebreak

\vskip15pt
\hfil\break
{\bf 4. Simple examples}

%\vskip5pt

\hfil\break
{\bf 4.1  HO+PPB case}

We shall here examine SMGT in a simple example that is represented 
by 1-dimensional harmonic oscillator (HO) +
1-dimensional parabolic potential barrier (PPB) 
$$V(x,y)=\HA m\omega^2x^2-\HA m\G^2 y^2, $$ 
where $m$ is a constant with the mass dimension. 
The eigenvalues of HO is well-known as 
\begin{equation}
\VE_{n_x}=\left( n_x+\HA\right) \hbar \omega 
\end{equation}
and 
the eigenvalues of PPB on the Gel'fand triplet are known 
to be pure 
imaginary values as [5--10]
\begin{equation}
\VE_{n_y}=\mp i\left( n_y+\HA\right) \hbar \G, 
\end{equation}
where $n_x$ and $n_y$ are natural numbers $n_x,\ n_y=0,1,2,\cdots$. 
It is known that the $\mp$ of the eigenvalues in PPB, 
respectively, stand for the decaying and growing 
resonance states. 
In this section we shall deal only with the states having 
the negative imaginary eigenvalues of PPB, 
which represent the decays of resonances for the time-scale 
$t\geqslant 0$ 
[1, 5--9]. 
Then the energy of a constituent is written by 
\begin{equation}
\VE_{n_xn_y}=\left(n_x+\HA\right)\hbar \omega
- i\left( n_y+\HA\right) \hbar \G.  
\end{equation}

\vskip5pt
\hfil\break
(1) Microcanonical ensemble

Let us start from microcanonical ensemble 
for the system composed of $N$ independent particles 
being in the above potential $V(x,y)$. 
The total complex energy of the system ${\cal E}$ is represented by 
\begin{equation}
{\cal E}_{M^\Re M^\Im}=\left(M^\Re+\HA N\right)\hbar \omega 
- i\left( M^\Im+\HA N\right) \hbar \G, 
\end{equation} 
where $M^\Re=\sum_{i=1}^N n_{xi}$ and $M^\Im=\sum_{i=1}^N n_{yi}$. 
Hereafter we shall use the notations $E=(M^\Re+ N/2)\hbar \omega$ 
for the total real energy and $\GA=(M^\Im+N/2)\hbar \G$ for the total 
imaginary energy. 
The thermodynamical weight %$W=M^\Re W^\Im$ 
is evaluated as 
\begin{equation}
W_N(M^\Re,M^\Im)=W_N^\Re(M^\Re) W_N^\Im(M^\Im), 
\end{equation} 
where 
\begin{eqnarray*}
W_N^\Re(M^\Re)&=&{(M^\Re+N-1)! \over M^\Re! (N-1)!}, \CR 
W_N^\Im(M^\Im)&=&{(M^\Im+N-1)! \over M^\Im! (N-1)!}.
\end{eqnarray*} 
The entropy is obtained by 
$$ %\begin{equation}
S({\cal E})= S^\Re(E) +S^\Im(\GA), 
$$ %\end{equation}
where the contributions from the real and imaginary parts are expressed 
in the same form as 
\begin{equation}
S^\bullet=k_{\rm B}\left[(M^\bullet+N) \log (M^\bullet+N)-
M^\bullet \log M^\bullet-
N\log N\right], 
\end{equation} 
where $\bullet$ denotes $\Re$ or $\Im$ and $M^\bullet,N\gg 1$ are postulated as 
usual. 
The complete symmetry between the contributions of HO and PPB 
in the entropy originates from the completely same structure 
of the total real and imaginary parts of the energy. 
We can introduce two temperatures corresponding to two constraints for 
giving the maximum of the entropy $S$ as [2] 
\begin{equation}
{1\over T}={\PA S^\Re \over \PA E}, \  \ \ 
{1\over T^\Im}={\PA S^\Im \over \PA \GA}. 
\end{equation} 
The explicit forms are obtained as 
\begin{equation}
{1\over T}={ k_{\rm B} \over \hbar \omega}\log {E/N+\hbar\omega/2 \over 
        E/N-\hbar\omega/2},\ \ \ 
{1\over T^\Im}={ k_{\rm B} \over \hbar \G}\log {\GA/N+\hbar\G/2 \over 
        \GA/N-\hbar\G/2}.
\end{equation}
Everything can be derived from the entropies, following the argument 
carried out in SMHS, e.g. %such that 
\begin{equation}
E=N\biggl(\HA \hbar \omega 
+{\hbar \omega \over \EX^{\B\hbar\omega}-1}\biggr),\ \ \ 
\GA=N\biggl(\HA \hbar \G +{\hbar \G \over \EX^{\B^\Im\hbar\G}-1}\biggr).
\end{equation} 
Since $\B^\Im=2t/\hbar$, 
we see the time-dependence of 
the total imaginary energy $\GA$ in the second equation of (29), 
which will be examined afterwards. 

\vskip5pt

\hfil\break
(2) Canonical ensemble

Following the argument given in $\S$ 2, the partition functions 
for the real and imaginary parts are obtained as 
\begin{equation}
Z_N^\bullet=\left({\EX^{- \B^\bullet \hbar \varOmega/2} \over 
1-\EX^{-\B^\bullet \hbar \varOmega}}\right)^N, 
\end{equation} 
where $\varOmega=\omega$ in the real part for $\bullet=\Re$ and 
$\varOmega=\G$ in the imaginary part for $\bullet=\Im$ should be taken. 
The derivations of the free energies $F^\Re$ and $F^\Im$ 
given in (7) are trivial. 
It is easy to examine that the mean values of $E$ and $\GA$ are  
same as those derived in (29) of microcanonical ensemble. 
The entropies of (9) are evaluated as 
\begin{equation}
S^\bullet=N k_{\rm B}
   \left[\B^\bullet \hbar \varOmega{\EX^{\B^\bullet \hbar \varOmega} \over 
   \EX^{\B^\bullet \hbar \varOmega}-1}-\log 
   (\EX^{\B^\bullet \hbar \varOmega}-1)\right].
\end{equation} 
We also 
easily see that they coincide with those given in (26) of microcanonical 
ensemble. 

\vskip5pt
\hfil\break
(3) Grand canonical ensemble

The present case is the independent particle model discussed in the last 
of the previous section. 
Then we can immediately get the partition function from (19);
\begin{equation}
\varXi={1 \over 1-\EX^{\B\mu}Z_1},
\end{equation} 
where the canonical partition function $Z_1$ for a particle is given by 
$$
Z_1=\biggl({\EX^{-\B \hbar \omega /2} \over 
    1-\EX^{-\B \hbar \omega}}\biggr)
    \biggl({\EX^{-\B^\Im \hbar \G /2} \over 
    1-\EX^{-\B^\Im \hbar \G}}\biggr). 
$$ 
The mean number is obtained as 
\begin{equation}
\bar N={\EX^{\B\mu}Z_1 \over 1-\EX^{\B\mu}Z_1}.
\end{equation} 
From this equation the chemical potential is expressed by  
\begin{equation}
\mu=\B^{-1}\left[\HA \B \hbar \omega +\log (1-\EX^{-\B\hbar\omega})+
            \HA \B^\Im \hbar \G +\log (1-\EX^{-\B^\Im\hbar\G})
            -\log (1+{1\over \bar N})\right].
\end{equation} 
For $\bar N \gg 1$ the contribution of 
the last term in the right-hand side of the above equation 
vanishes.
Then we see the behavior of $\mu$ for small $t$ as follows;
\begin{equation}
\mu \sim \log \G t \ \ \ \text{for}\ \ t\rightarrow 0. 
\end{equation} 
The divergence at $t=0$ appears so as to cancel the divergence of 
$Z_1^\Im$ at 
$t=0$, because in the canonical distribution (4) the dumping factor 
$\EX^{-\B^\Im \GA}$ disappears at $t=0$ 
and then $Z_1^\Im$ becomes infinity, 
of which divergence is easily obtained as the $t^{-1}$ type. 
Note here that the divergences also appear in $S^\Im$ and 
$\bar \GA$ as $\log t$ and $t^{-1}$ types, respectively. 

Thus we obtain the $t$-dependence of all thermodynamical quantities for 
the systems involving unstable states for small $t$-values. 

\pagebreak

\vskip5pt
\hfil\break
{\bf 4.2 $\boldsymbol{d}$-dimensional free motion + PPB case}

Let us briefly discuss one more example described by the $d$-dimensional 
free motion + PPB, where the equation of states 
with respect to the temperature $T$, volume $V$ and pressure $p$ are 
treatable. 
Here we study the problem in terms of $T$-$p$ distribution of which 
partition function is defined by 
\begin{equation}
Y=\int_0^\infty \EX^{-\B pV} Z_NdV, 
\end{equation} 
where the canonical partion function $Z_N=Z_N^\Re Z_N^\Im$. 
The real part $Z_N^\Re$ for the free motions is given by 
\begin{equation}
Z_N^\Re={1 \over N!}{1 \over (2\pi \hbar )^{dN} }V^N 
(2\pi m k_{\rm B}T)^{dN/2}
\end{equation} 
and the imaginary part $Z_N^\Im$ for the 1-dimensional PPB 
is the same as that of the previous model. 
After the integration we have 
\begin{equation}
Y={1 \over (2\pi \hbar )^{dN}}(2\pi m k_{\rm B}T)^{dN/2}
   \left({k_{\rm B}T \over p}\right)^{N+1}Z_N^\Im.
\end{equation}   
From the thermodynamical relation 
%\begin{equation}
$G=-\B^{-1} \log Y$ 
for $N \gg 1$ 
we obtain 
\begin{equation}
G=-N\B^{-1}\left[{d+2 \over 2}\log T-\log p+\log {m^{d/2}
k_{\rm B}^{(d+2)/2} \over (2\pi \hbar^2 )^{d/2}}-
\HA\B^\Im \hbar \G - \log (1-\EX^{-\B^\Im \hbar \G})\right].
\end{equation} 
The equation of states is immediately derived from the relation 
$V=\partial G/\partial p$ as usual 
$$
pV=Nk_{\rm B} T. 
$$ 
Note here that this equation describes the relation between $V$ and $p$ 
for the free motions. 
In order to answer the question whether physical quantities for the 
imaginary freedom 
corresponding to the volume and the pressure are meaningful or not, 
we have to study the meanings of continuous imaginary spectra on 
Gel'fand triplet, which 
do not represent usual resonances described by the Breit-Wigner 
resonance formula in cross-sections. 

The chemical potential is gotten  from the relation 
$ G=\mu N$ as 
\begin{equation}
\mu=k_{\rm B} T\left[\log {p \over k_{\rm B} T} 
\left({2\pi \hbar^2 \over m k_{\rm B} T}\right)^{d \over 2}  
+ \HA\B^\Im \hbar \G + \log (1-\EX^{-\B^\Im \hbar \G})\right].
\end{equation} 
It has the $t$-dependence of the $\log t$ type at small $t$, 
which is same as the previous case given by (34). 
The same result for $\mu$ can be obtained in grand canonical ensemble, 
where the number $N$ should be replaced by the mean number $\bar N$. 

%\vskip5pt

From the above examples we see that SMGT is applicable to realistic processes. 

\vskip15pt
\hfil\break
{\bf 5. Entropy transfer from $S^\Im$ to $S^\Re$}
\vskip5pt

Let us consider the entropy transfer from $S^\Im$ to $S^\Re$ 
in an adiabatic process described by a decay of a system 
that is composed of $N$ resonances in a 1-dimensional PPB+some ordinary 
potentials, where the ordinary potentials mean potentials which are 
described by Hilbert spaces and the systems described the potentials 
can have thermal equilibriums of SMHS. 
We can, therefore, consider that $S^\Im$ and $S^\Re$, respectively, 
stand for the entropy of the PPB system and that of the ordinary system. 
Here we study the process where the decays of the resonance 
system are absorbed into the system described by the ordinary potentials. 
After the decay processes are opened at $t=0$, 
the entropy of the system being in the PPB is obtained from (31) as 
\begin{equation}
S^\Im=N k_{\rm B}
\left[2\G t{e^{2\G t} \over  e^{2\G t}-1}-\log (e^{2\G t}-1)\right].
\end{equation} 
For small $t$ such that 
$\G t\ll 1/2$ 
the entropy behaves 
\begin{equation}
S^\Im\simeq -N k_{\rm B}\log\tau 
\end{equation} 
where $\tau=\G t$. 
As already noted, it diverges at $t=0$.
This relation gives us 
\begin{equation}
dS^\Im=-N k_{\rm B}{d\tau \over \tau} \ \ \ \text{for}\ \tau\ll\HA. 
\end{equation} 
Since the total entropy conserves in the adiabatic process, that is, 
the relation 
\begin{equation}
dS=dS^\Re+dS^\Im=0
\end{equation} 
holds, we have the relation 
\begin{equation}
dS^\Re=-dS^\Im. 
\end{equation} 
Note here that $dS^\Re$ is always positive because $dS^\Im<0$ is kept. 
In the system described only by PPB's 
the temperature $T$ originated from the freedom of 
real energy eigenvalues is zero, i.e. $T=0$, 
since the system has no real energy freedom. 
This means that the temperature must be zero at $t=0$, i.e., 
just at the moment when the decay processes 
are opened. 
Let us write it as 
\begin{equation}
T(t)=K_0\tau^\delta \ \ \ \text{for}\ \tau\ll\HA. 
\end{equation} 
where  $K_0$ and $\delta$ should be  positive constants. 
Since the direct observable in this process is 
the real energy $E^\Re$ 
released into the ordinary potentials by the decay of resonances, 
we should evaluate the real energy produced  
in this process. 
For the small $t$ we have 
\begin{equation}
dE^\Re=T(t)dS^\Re=N k_{\rm B}K_0\tau^{\delta-1}d\tau 
\ \ \ \text{for}\ \tau\ll\HA. 
\end{equation} 
Then we can estimate the real energy produced in the process 
during the short period from $0$ to $t$ ($\ll 1/2\G$) as 
\begin{equation}
E^\Re=\int_0^{\G t} {dE^\Re \over d\tau}d\tau
        =N k_{\rm B} {K_0 \over \delta} (\G t)^\delta. 
\end{equation} 
Since $\delta>0$, this process produces a real  positive energy. 
Even if $S^\Im$ diverges at $t=0$, we obtain a finite energy production. 
The unknown constants $K_0$ and $\delta$ will depend on  
the property of the system where the produced energy is 
absorbed. 
We see that the system in PPB's can be the source of the energy production. 
It, of course, does not mean the break down of the energy conservation 
law. 
In the process where the system is composed in the PPB
the real energy produced in the decay process is stored 
as $S^\Im$ in the system. 
This means that the total produced energy which is evaluated by 
the integration from $t=0$ to $\infty$ must coincide with the energy 
consumed in the process for making the initial system. 
This integration will derive a relation between $K_0$ and $\delta$. 

\vskip15pt 
\hfil\break
{\bf 6. Equilibrium of a process described 
by reactions $A+CB\rightleftarrows AC+B$}

\vskip5pt

As discussed by Child, the chemical reaction $A+CB\rightarrow AC+B$ 
is well described by the potential having two bumps [11, 12]. 
Connor studied the reaction by representing the potential 
in terms of PPB's [13]. 
They investigated the reaction cross-sections of the processes 
in the WKB method and showed that the cross-sections were given by 
the Breit-Wigner resonance formula. 
The Breit-Wigner formulas of the cross-sections for PPB scatterings 
have already been 
obtained in our scheme based on the Gel'fand triplet [9]. 
Here we shall study chemical equilibriums of the systems 
containing two reactions $A+CB\rightarrow AC+B$ and 
$AC+B\rightarrow A+CB$ ($A+CB\rightleftarrows AC+B$) simultaneously. 
We study the case where the potentials for the exchanged particle $C$ is 
described by two 1-dimensional PPB's 
having the centers at the positions of $A$ and $B$ 
which are spatially separated enough to treat them 
as two independent systems. 
The PPB constants of the systems $A$ and $B$ are denoted by 
$\G_1$ and $\G_2$, respectively. 
The total systems are described by an ensemble 
composed of $N$-number of independent reactions 
$A+CB\rightleftarrows AC+B$.   
In the present discussion we postulate that the systems $A$ and $B$ are 
heavy enough to neglect their movements in the 
interactions with $C$. 
Note here that the reaction $A+CB\rightarrow AC+B$  describing the process 
that the particle $C$ is approaching to $A$ 
is understood as the growing resonance state for the system $A$, 
but the same process is, on the other hand, understood 
as the decaying resonance state for the system $B$ 
because the particle $C$ is leaving from $B$. 
The reaction $AC+B\rightarrow A+CB$ describing 
the process that $C$ is approaching to $B$ is understood  
vice versa.  
From this consideration on the growing and decaying resonance 
states, we see that there are the following relations between 
the number of the growing resonances $N_1^-$ for the system $AC$ 
and that of the decaying resonances $N_2^+$ for the system $BC$ and also 
between the number of the decaying ones $N_1^+$ for $AC$ and that 
of the growing ones $N_2^-$ for $BC$; 
\begin{equation}
N_1^-=N_2^+,\ \ \ N_1^+=N_2^-, 
\end{equation} 
Thus the total number $N$ is expressed by the sum 
$N=N_1^- +N_1^+$ provided that we pay attention to the system $AC$, 
whereas it is written down by 
the sum  $N=N_2^+ +N_2^-$ from the side of the system $BC$. 
In microcanonical ensemble the imaginary parts of the energies 
of the growing and decaying 
states for the reactions $A+CB\rightleftarrows AC+B$ are, respectively, 
given by 
\begin{equation}%$$
\left.
\begin{aligned}
\GA_1^-=\left(M_1^- +\HA N_1^-\right)\hbar\G_1,\ \ 
\GA_2^+=\left(M_2^+ +\HA N_1^-\right)\hbar\G_2, 
\ \ &\text{for }& A+CB\rightarrow AC+B,  \\ %\CR
%$$ 
%$$
\GA_2^-=\left(M_2^- +\HA N_1^+\right)\hbar\G_2,\ \ 
\GA_1^+=\left(M_1^+ +\HA N_1^+\right)\hbar\G_1,
\ \ &\text{for}& AC+B\rightarrow A+CB, 
\end{aligned} \right\} 
\end{equation} %$$
where the imaginary parts are defined by 
${\cal E}_i^\pm=\mp\I \GA_i^\pm$ (suffix $i=1,2$), 
$M_i^\pm=0,1,2,\cdots$ and the relations of (49) are used. 
Note that 
the total imaginary energies of $AC$ and $BC$ are written by 
${\cal E}_1=-\I(\GA_1^+ -\GA_1^-)$ and 
${\cal E}_2=-\I(\GA_2^+ -\GA_2^-)$, respectively. 
In the equilibrium the relations 
\begin{equation}
\GA_1^-=\GA_2^+,\ \  \ \GA_1^+=\GA_2^- 
\end{equation} 
must be satisfied, since the $t$-dependence of the canonical ensemble 
for the reaction $A+CB\rightarrow AC+B$, which is given by 
$\EX^{-\B^\Im(\GA_2^+ -\GA_1^-)}$, and that for $AC+B\rightarrow A+CB$ 
given by $\EX^{-\B^\Im (\GA_1^+ -\GA_2^-)}$ must vanish in the equilibrium. 
Now we have the thermodynamical weight as 
\begin{equation}
W=W_1W_2,
\end{equation} 
where 
\begin{equation}%\begin{eqnarray} 
\left.
\begin{aligned}
W_1&=&{(M_1^++N_1^+-1)! \over M_1^+!(N_1^+-1)!}
    {(M_1^-+N-N_1^+-1)! \over M_1^-!(N-N_1^+-1)!}, \\%\CR
W_2&=&{(M_2^-+N_1^+-1)! \over M_2^-!(N_1^+-1)!}
    {(M_2^++N-N_1^+-1)! \over M_2^+!(N-N_1^+-1)!}. 
\end{aligned} \right\} 
\end{equation}%\end{eqnarray} 
The maximum of the entropy is realized at the point, where the relation 
\begin{equation}
{\partial \over \partial N_1^+}\log W =
\log {(M_1^++N_1^+)(M_2^-+N_1^+)(N-N_1^+)^2 \over 
          (M_1^-+N-N_1^+)(M_2^++N-N_1^+)(N_1^+)^2}=0, 
\end{equation} 
is fulfilled, where $M,N\gg 1$ are used. 
We have the equation satisfied in the equilibrium 
\begin{equation}
(M_1^-+N_1^-)(M_2^++N_1^-)(N_1^+)^2=(M_1^++N_1^+)(M_2^-+N_1^+)(N_1^-)^2,
\end{equation} 
where $N_1^-=N-N_1^+$ is put. 
By using the relations of (50) and (51) 
we obtain the equation 
\begin{equation} 
\left({\GA_1^-\over N_1^-}+\HA\hbar\G_1 \right)
\left({\GA_1^-\over N_1^-}+\HA\hbar\G_2 \right)=
\left({\GA_1^+\over N_1^+}+\HA\hbar\G_1 \right)
\left({\GA_1^+\over N_1^+}+\HA\hbar\G_2 \right).
\end{equation} 
Taking account of the constraints 
$\GA_1^-/N_1^->0$ and $\GA_1^+/N_1^+>0$, we get the solution 
\begin{equation}
\frac{\GA_1^-}{N_1^-}=\frac{\GA_1^+}{N_1^+}.
\end{equation} 
This result shows that the mean grow width for a growing resonance, 
$\GA_1^-/ N_1^-$, is equal to the mean decay width for a decaying resonance, 
$\GA_1^+/ N_1^+$, for the system $AC$. 
From the relations of (49) and (51) we can, of course, derive the relation 
$\GA_2^-/ N_2^-=\GA_2^+/ N_2^+$ for the system $BC$. 
Generally the relations 
\begin{equation}
\frac{\GA_1^-}{N_1^-}=\frac{\GA_1^+}{N_1^+}=
\frac{\GA_2^-}{N_2^-}=\frac{\GA_2^+}{N_2^+}
\end{equation} 
are obtained. 
These relations indicate that a kind of balance 
like a detailed balance is held 
between the grow processes and the decay ones 
in the reaction $A+CB\rightleftarrows AC+B$. 

Though this model is too much simple to describe realistic chemical processes, 
we can at least say that this scheme (SMGT) is consistent with 
our primitive understandings. 

\vskip15pt
\hfil\break
{\bf 7. Discussions}
\vskip5pt

We have proposed a statistical mechanics which can contains unstable 
states on Gel'fand triplets (SMGT) and applied it to  a few simple 
processes. 
The validity of this theoretical scheme will be examined 
by applying it to many realistic processes and by comparing 
with experiments. 
We should, however, remember that SMGT is applicable to the processes 
where the change of systems with respect to time are so slow  that 
the systems can be dealt with as being 
in a thermal equilibrium at any moment. 

Here we shall comment on a general formula for the equation of motion 
for the mean values in canonical ensembles. 
Provided that the real and imaginary parts are separable as the canonical 
distribution given by (4), 
the mean value of the quantity $A(\GA)$ is obtained by 
\begin{equation}
\bar A=\left.\int A(\GA) \EX^{-\B^\Im \GA} W^\Im(\GA) d\GA\right/
        \int  \EX^{-\B^\Im \GA} W^\Im(\GA) d\GA. 
\end{equation} 
In general we should consider that the average with respect to the real energy 
part has already been taken as for $A(\GA)$.  
The derivative of $\bar A$ with respect to $t$ is evaluated as 
\begin{eqnarray}
{d \bar A \over d t}&=&{2 \over \hbar} \left\{
{-\int \GA A(\GA) \EX^{-\B^\Im \GA} W^\Im(\GA) d\GA \over 
   \int  \EX^{-\B^\Im \GA} W^\Im(\GA) d\GA} +
   {\int A(\GA) \EX^{-\B^\Im \GA} W^\Im(\GA) d\GA
    \int \GA \EX^{-\B^\Im \GA} W^\Im(\GA) d\GA
  \over [\int \EX^{-\B^\Im \GA} W^\Im(\GA) d\GA]^2}\right\}  \CR
  &=&{2 \over \hbar}\left(-\overline{\GA A} +\bar \GA \bar A\right).
  \end{eqnarray} 
For $A=\GA$ we have the equation 
\begin{equation}
{d\bar \GA \over dt}=-{2 \over \hbar} (\Delta \GA)^2 <0, 
\end{equation} 
where $(\Delta \GA)^2={\bar \GA^2}-(\bar \GA)^2$. 
This equation means that $\bar \GA$ becomes small in the time evolution 
in all processes. 
Considering the fact that states with large imaginary 
eigenvalues decay rapidly, we can comply with this result. 

Throughout this paper we have discussed the cases where the total 
real and imaginary parts $E$ and $\GA$ are independently determined. 
Gel'fand triplets, however, contain many other solutions 
such that the real and imaginary eigenvalues $\E$ and $\G$ 
have some correlations. %[13]. 
In such processes the thermodynamical weight cannot be obtained 
by the simple product of $W^\Re$ and $W^\Im$ as given in (2) [2]. 
Study of such processes is still an open question in  the present SMGT. 

%\pagebreak

\vskip15pt
\hfil\break
{\bf References}%{\Large References}}
\vskip5pt
  \hfil\break
  [1] A. Bohm and M. Gedella, {\it Dirac Kets, Gamow Vectors and 
  Gel'fand Triplets}, Lecture Notes in Physics {\bf 348} 
  (Springer, Berlin) 1989. 
  \hfil\break
  [2] T. Kobayashi and T. Shimbori, 
  {\it Statistical Mechanics for States with Complex Eigenvalues 
  and Quasi-stable Semiclassical Systems}, cond-mat/0005237. 
  \hfil\break
  [3] T. Kobayashi and T. Shimbori, 
  {\it Entropy Burst from Parabolic Potentials}, cond-mat/ 0007014. 
  \hfil\break
  [4] T. Shimbori and T. Kobayashi, 
  {\it Stationary Flows of the Parabolic Potential Barrier in Two Dimensions},
   quant-ph/0006019. 
  \hfil\break
  [5] G. Barton, Ann. Phys. {\bf 166} (1986) 322.
  \hfil\break
  [6] P. Briet, J. M. Combes and P. Duclos, Comm. Partial Differential 
  Equations {\bf 12} (1987) 201.
  \hfil\break
  [7] N. L. Balazs and A. Voros, Ann. Phys. {\bf 199} (1990) 123.
  \hfil\break
  [8] M. Castagnino, R. Diener, L. Lara and G. Puccini, Int. J. Theor. Phys. 
  {\bf 36} (1997) 2349. % quant-ph/0006011.
  \hfil\break
  [9] T. Shimbori and T. Kobayashi, Nuovo Cim. {\bf 115B} (2000) 325. 
  %math-ph/9910009. 
  \hfil\break
  [10] T. Shimbori, Phys. Lett. {\bf A273} (2000) 37. 
  \hfil\break
  [11] M. S. Child, Proc. Roy. Soc. (London) {\bf A292} (1966) 272. 
  \hfil\break
  [12]  M. S. Child, Mol. Phys. {\bf 12} (1967) 401.
  \hfil\break
  [13]  J. N. L. Connor, Mol. Phys. {\bf 15} (1968) 37. 
\end{document}